\documentclass[pra, 12pt]{revtex4}
\usepackage{graphics}

\newcommand {\psig}{\Sigma_}
\newcommand {\pxi}{\Xi_}

\newcommand {\Tr}{{\mbox{Tr}}}

\begin{document}
\title{Repeatable procedures and maps in open quantum dynamics}
\author{Thomas F. Jordan}
\email[email: ]{tjordan@d.umn.edu}
\affiliation{Physics Department, University of Minnesota, Duluth, Minnesota 55812}
\author{Anil Shaji}
\email[email: ]{shaji@unm.edu}
\affiliation{Department of Physics and Astronomy, The University of New Mexico, 800 Yale Boulevard NE, Albuquerque, New Mexico 87131, USA} 

\begin{abstract}
Examples of repeatable procedures and maps are found in the open quantum dynamics of one qubit that interacts with another qubit. They show that a mathematical map that is repeatable can be made by a physical procedure that is not.
\end{abstract}


\maketitle

The mathematics of maps in open quantum dynamics does not always conform to the physics. The divergence can be wide in the case of inverses. We consider the open quantum dynamics of a subsystem made by dynamics described by a unitary operator in a larger system. The change of states of the subsystem is described by a trace-preserving completely-positive linear map \cite{NielsenChuang} when there are no initial correlations between the subsystem and the rest of the larger system. It is common for a trace-preserving completely-positive map to have an inverse, but the inverse is generally not completely positive. Every trace-preserving completely-positive linear map comes from dynamics described by a unitary operator for a larger system, but the inverse map generally does not describe the reversed dynamics. A trace-preserving completely-positive map that is unital can not have an inverse that is obtained from any dynamics described by any unitary operator for any larger system, even when maps that are not completely positive are allowed to describe the change of states of the subsystem when there are correlations between the subsystem and the rest of the larger system \cite{Inverses}.

Here we will consider another case, the recently introduced concept of repeatable maps \cite{RybarZiman}. We will look at examples in the open quantum dynamics of one qubit interacting with another qubit. There we will see a repeatable mathematical map made by a physical procedure that is not repeatable.

Consider two quantum systems $S$ and $R$ and an interaction between them described by a unitary operator $U$ for the system of $S$ and $R$ combined. Suppose the initial state is described by a product of density matrices $\rho $ for $S$ and $\xi $ for $R$. Then after the interaction the states of $S$ and $R$ are described by the density matrices 
\begin{eqnarray}
  \label{Uproc}
  \lfloor R,U,\xi \rfloor (\rho ) & = & \Tr_R[ U\rho \xi U^{\dagger}] \nonumber \\
  \lceil S,U,\rho \rceil (\xi ) & = & \Tr_S[ U\rho \xi U^{\dagger}].
\end{eqnarray}
The density matrix $\rho $ for $S$ is changed by a procedure $\lfloor R,U,\xi \rfloor $ that depends on $R$, $U$, and $\xi $, and the density matrix $\xi $ for $R$ is changed by a similar procedure $\lceil S,U,\rho \rceil $ that depends on $S$, $U$, and $\rho $. We will write just $\lfloor \xi \rfloor $ and $\lceil \rho \rceil $ for $\lfloor R,U,\xi \rfloor $ and $\lceil S,U,\rho \rceil $. We say that the procedure $\lfloor \xi \rfloor $ is \textit{repeatable} \cite{RybarZiman} if
\begin{equation}
\label{repeatabledef}
\lfloor \lceil \rho \rceil (\xi )\rfloor (\rho' ) = \lfloor \xi \rfloor (\rho ')
\end{equation}
for all density matrices $\rho $ and $\rho'$ for $S$. The procedure gives the same result when it is repeated without $\lceil \rho \rceil (\xi )$ being reset to $\xi $. This is implied by, but is weaker than, the condition that the state of the system $S$ is changed without any accompanying change in the state of its environment $R$. That is the condition under which the Markov approximation holds and we can describe the open evolution of $S$ with a master equation of the Kossakowski-Lindblad type~\cite{kossakowski72a,lindblad76a}.

The map of density matrices $\rho $ defines a trace-preserving completely-positive linear map \cite{NielsenChuang} of matrices for $S$. The map is called \textit{repeatable} \cite{RybarZiman} if it is made by a repeatable procedure. The same map can be made by different procedures involving different systems $R$ with states represented by different density matrices $\xi $ and interactions described by different unitary operators $U$. A map that is repeatable can be made by a procedure that is not; this means the map is also made by a procedure that is repeatable.

A repeatable procedure may be more useful than a repeatable map. When a repeatable procedure is used to change the state of $S$, it can be used to change the state of $S$ again, or to change the state of another system identical to $S$, without changing $R$ or $U$ or resetting the state of $R$. When the state of $S$ is changed by a repeatable map made by a procedure that is not repeatable, the map can not be used again without changing or resetting the procedure. We will see from our examples that when conditions imply that a map must be repeatable, a procedure that makes the map may still be not repeatable.

Rybar and Ziman \cite{RybarZiman} showed that if a map is a mixture of unitary maps, which means it changes each density matrix $\rho $ for $S$ to $\sum_k p_k U_k \rho U_k^\dagger $ where the $U_k$ are unitary operators for $S$ and the $p_k$ are positive numbers whose sum is $1$, then the map is repeatable an infinite number of times. They also show that if a trace-preserving completely-positive map is repeatable an infinite number of times, it must be unital, which means it does not change the unit matrix for $S$. They observe that for a qubit every unital trace-preserving completely-positive map is a mixture of unitary maps, so for a qubit a trace-preserving completely-positive map is repeatable an infinite number of times if and only if it is unital. We will see from our examples that a trace-preserving completely-positive unital map for a qubit can be made by a procedure that is not repeatable.

For the examples, consider two qubits $S$ and $R$ described by Pauli matrices $\psig 1$, $\psig 2$, $\psig 3$ for $S$ and $\pxi 1$, $\pxi 2$, $\pxi 3$ for $R$. The states of $S$ and $R$ can be described by the mean values $\langle \psig j \rangle $ for $j = 1,2,3$ and $\langle \pxi k \rangle $ for $k =1,2,3$ which are changed to $\langle U^\dagger \psig j U\rangle $ and $\langle U^\dagger \pxi k U\rangle $. The density matrices
\begin{equation}
\label{rhosum}
\rho  = \frac{1}{2}(\openone  + \sum_{j=1}^3 \langle \psig j \rangle \psig j ) 
\end{equation} 
and
\begin{equation}
\label{xisum}
\xi  = \frac{1}{2}(\openone  + \sum_{k=1}^3 \langle \pxi k \rangle \pxi k )
\end{equation} 
for $S$ and $R$ are changed to 
\begin{equation}
\label{rhoUsum}
\lfloor \xi \rfloor (\rho ) = \frac{1}{2}(\openone  + \sum_{j=1}^3 \langle U^\dagger \psig j U\rangle \psig j ) 
\end{equation} 
and
\begin{equation}
\label{xiUsum}
\lceil \rho \rceil (\xi ) = \frac{1}{2}(\openone  + \sum_{k=1}^3 \langle U^\dagger \pxi k U\rangle \pxi k ).
\end{equation}
Let
\begin{equation}
  \label{Udef}
  U = e^{-i \frac{1}{2} \left(\gamma_2 \psig 2 \pxi 2 + \gamma_3 \psig 3 \pxi 3 \right)} .
\end{equation}
It is easy to calculate \cite{me79} that
\begin{eqnarray}
  \label{Umv}
  \langle U^{\dagger}\psig 1 U\rangle & = & \langle \psig 1 \rangle \cos \gamma_2 \cos \gamma_3 + \langle \pxi 1 \rangle \sin \gamma_2 \sin \gamma_3 \nonumber \\
 &&  - \langle \psig 2 \rangle \langle \pxi 3 \rangle \cos \gamma_2 \sin \gamma_3 + \langle \psig 3 \rangle \langle \pxi 2 \rangle \sin \gamma_2 \cos \gamma_3 , \nonumber \\
  \langle U^{\dagger}\psig 2 U \rangle &=& \langle \psig 2 \rangle \cos \gamma_3 + \langle \psig 1 \rangle \langle \pxi 3 \rangle \sin \gamma_3 , \nonumber \\
  \langle U^{\dagger}\psig 3 U\rangle   &=& \langle \psig 3 \rangle \cos \gamma_2  - \langle \psig 1 \rangle \langle \pxi 2 \rangle \sin \gamma_2 . 
\end{eqnarray}
Here $\langle \psig j \pxi k \rangle  = \langle \psig j \rangle \langle \pxi k \rangle $ in the initial state. These equations (\ref{Umv}) are not changed when the two qubits are interchanged. 

If $\gamma_2$ is zero, then
\begin{eqnarray}
  \label{Umvr}
  \langle U^{\dagger}\psig 1 U\rangle & = & \langle \psig 1 \rangle \cos \gamma_3  - \langle \psig 2 \rangle \langle \pxi 3 \rangle \sin \gamma_3 , \nonumber \\
  \langle U^{\dagger}\psig 2 U \rangle &=& \langle \psig 2 \rangle \cos \gamma_3 + \langle \psig 1 \rangle \langle \pxi 3 \rangle \sin \gamma_3 , \nonumber \\
  \langle U^{\dagger}\psig 3 U\rangle   &=& \langle \psig 3 \rangle , \nonumber \\
\langle U^{\dagger}\pxi 3  U\rangle   &=& \langle \pxi 3  \rangle . 
\end{eqnarray}
Then the procedure is repeatable. It depends on only $\gamma_3 $ and $\langle \pxi 3 \rangle $, and they are not changed.

If neither $\gamma_2$ nor $\gamma_3$ is zero, the procedure is not repeatable. It depends on $\langle \pxi 1 \rangle $, $\langle \pxi 2 \rangle $, and $\langle \pxi 3 \rangle $, which are changed for most values of $\langle \psig 1 \rangle $, $\langle \psig 2 \rangle $, $\langle \psig 3 \rangle $. If $\langle \pxi 1 \rangle $ is zero, the map is unital. Then the map is repeatable, because it is a trace-preserving completely-positive map and for a qubit every unital trace-preserving completely-positive map is repeatable. This is an example of a repeatable map made by a procedure that is not repeatable.

Here are examples of repeatable procedures that make this repeatable map in the cases where $\langle \pxi 1 \rangle $ is zero and $|\langle \pxi 2 \rangle | + |\langle \pxi 3 \rangle |$ is $1$. They use new systems $R$ that are larger than a single qubit. Let
\begin{equation}
  \label{Usum}
  U = \sum_j U_j E_j
\end{equation}
with the $U_j $ unitary operators for $S$ and $R$ combined and the $E_j $ projection operators for $R$ such that all the $U_j $ commute with all the $E_k $ and
\begin{equation}
  \label{Ejprop}
E_j E_k = \delta_{jk} E_j, \quad \sum_j E_j = 1.
\end{equation}
Then $U$ is a unitary operator. Let
\begin{eqnarray}
  \label{U1234}
U_1 & = &  e^{-i \frac{1}{2} \gamma_2 \psig 2 \Upsilon_2 } e^{-i \frac{1}{2} \gamma_3 \psig 3 \Pi_3 }, \nonumber \\
U_2 & = &  e^{i \frac{1}{2} \gamma_2 \psig 2 \Upsilon_2 } e^{-i \frac{1}{2} \gamma_3 \psig 3 \Pi_3 }, \nonumber \\
U_3 & = &  e^{-i \frac{1}{2} \gamma_3 \psig 3 \Pi_3 } e^{-i \frac{1}{2} \gamma_2 \psig 2 \Upsilon_2 }, \nonumber \\
U_4 & = & e^{i \frac{1}{2} \gamma_3 \psig 3 \Pi_3 } e^{-i \frac{1}{2} \gamma_2 \psig 2 \Upsilon_2 } 
\end{eqnarray}
with $\Upsilon_2 $ and $\Pi_3 $ Pauli matrices for two qubits that are part of $R$. Other $U_j $ in the sum (\ref{Usum}) will not be used. Let $E_1 $, $E_2 $, $E_3 $, $E_4 $ project onto subspaces for $R$ where the vectors are eigenvectors of $\Upsilon_2 $ and $\Pi_3 $ with eigenvalues $1$ or $-1$ that are the sign of $\langle \pxi 2 \rangle $ for $\Upsilon_2 $ and the sign of $\langle \pxi 3 \rangle $ for $\Pi_3 $. Let
\begin{equation}
  \label{stateR12}
\langle E_1 \rangle  = \langle E_2 \rangle  = \frac{1}{2}|\langle \Xi_3 \rangle |,
\end{equation}
\begin{equation}
  \label{stateR34}
\langle E_3 \rangle  = \langle E_4 \rangle  = \frac{1}{2}|\langle \Xi_2 \rangle |,
\end{equation}
\begin{equation}
  \label{stateRsum}
\langle E_1 \rangle  + \langle E_2 \rangle + \langle E_3 \rangle  + \langle E_4 \rangle  = 1
\end{equation}
for the state of $R$. Then $\langle E_j \rangle $ is zero for the other $E_j $ in the sum (\ref{Usum}), so they will not contribute to the result. This describes procedures that make the map described by Eqs.(\ref{Umv}) in the cases where $\langle \pxi 1 \rangle $ is zero and $|\langle \pxi 2 \rangle | + |\langle \pxi 3 \rangle |$ is $1$; this can be checked by calculating the $\langle U^\dagger \Sigma_j U\rangle $. These procedures are repeatable because they do not change the $E_j $ and do not change the state of $R$. The original procedure where $R$ is one qubit is not repeatable, but the new procedures with larger systems $R$ give the same map and are repeatable.

\section*{Acknowledgements}

Anil Shaji acknowledges the support of the NSF through contract No.~PHY-0803371 and the Office of Naval Research through Grant No.~N00014-07-1-0304.

\bibliography{RepeatBib}
\end{document}